# An exact and O(1) time heaviest and lightest hitters algorithm for sliding-window data streams

Remous-Aris Koutsiamanis    Pavlos S. Efraimidis

Department of Electrical and Computer Engineering,
Democritus University of Thrace,
University Campus, 67100 Xanthi, Greece
`{akoutsia,pefraimi}@ee.duth.gr`

**Abstract.** In this work we focus on the problem of finding the heaviest-$k$ and lightest-$k$ hitters in a sliding window data stream. The most recent research endeavours [6] have yielded an $\epsilon$-approximate algorithm with update operations in constant time with high probability and $O(1/\epsilon)$ query time for the heaviest hitters case. We propose a novel algorithm which for the first time, to our knowledge, provides *exact*, not approximate, results while at the same time achieves O(1) time with high probability complexity on both update and query operations. Furthermore, our algorithm is able to provide both the heaviest-$k$ and the lightest-$k$ hitters at the same time without any overhead. In this work, we describe the algorithm and the accompanying data structure that supports it and perform quantitative experiments with synthetic data to verify our theoretical predictions.

## 1 Introduction

The problem of finding the heaviest hitters, in its simplest form, is the problem of finding which category of items in a long succession of them is the most frequent one. This problem has been studied extensively in the last decade. The main reason for this is that a number of applications, some of them quite pervasive, need to solve it to provide enhanced services. The first application, and the one that mainly motivated this work, is network traffic monitoring (and shaping) on Internet routers. Being able to tell at any moment in time which set of packets is the most frequent passing through a router helps in both being able to tell what may be causing problems and subsequently resolving these problem in a "fair" manner towards those not contributing to the problem. Another application is financial data streams, where it is useful, for example, to know which stocks are showing the most mobility. Other applications include sensor networks, behaviour analysis on websites and trend tracking of hot topics.

The problem was first posed by Moore in 1980 and together with Boyer they presented the solution (in [3]) for finding the majority hitter in the basic version of the problem, i.e., non-window-based data streams. The basic heaviest hitters problem consists of a data stream where at each moment in time one item, which belongs to some itemset, arrives for processing. The goal is to be able to provide a list of the itemsets whose item counts are above a given $\theta$ threshold. Given the unbounded number of itemsets and length of the data stream, this cannot be achieved without unbounded memory. As a result, all of the proposed solutions for this problem have provided approximate results. This problem was studied and approximate solutions were proposed much later and concurrently by [4, 7].

Since, a significant body of work has been performed on both the basic problem and on its numerous variations. A good presentation of this work can be found in [8, 9]. The variant of the basic problem addressed in this work stems from the observation that only a section of the whole history of the data stream may be interesting. Usually, the most recent items are considered to be more important. This is one of the most common and arguably one of the most useful of these variations: finding the heaviest (and lightest) hitters in a sliding-window data stream.

In the sliding window model at each moment in time a constant number of items participate in a window over the data stream. This window always contains the most recent $Q$ items. This scenario resembles the operation of a queue with an upper limit on its capacity. As items arrive to be processed they are inserted at the end of the queue and as items are processed they are removed from the front of the queue.

All the algorithms proposed for both the basic problem and the sliding window variation have in common the requirement that they be able to operate on-line. This entails being able to do only one pass over the data, i.e., each arriving item may be examined only once by the algorithm. This is usually called an update operation and the complexity of this operation must be constant time. Furthermore, querying for the heaviest hitters must also be as fast as possible, ideally proportional to the number $k$ of the heaviest or lightest hitters that we request to be found.

The novelty of our algorithm is twofold, featuring for the first time, to our knowledge, the ability:

1. To provide *exact* results in the query operation and at the same time maintain constant time update and query operations.
2. To provide not only the heaviest but also the lightest hitters in the sliding window with the same performance and no overhead.

In the following sections we will describe the HL-HITTERS abstract data type which allows us to solve this problem. Moving on to the implementation, we describe the building blocks which we use to construct the data structure and some of their characteristics. We then describe the data structure itself and the algorithms which implement the HL-HITTERS operations. Subsequently, we present an experimental evaluation of the proposed solution and discuss its results. Finally, we propose some interesting possible extensions to this work.

## 2 Abstract Data Type

In order to provide an accurate description of our algorithm and the accompanying data structure we describe here its interface. The abstract data type which we define supports the operations shown in Table 1. All the operations in our implementation have constant time complexity.

| Operation | Input | Output | Description |
|---|---|---|---|
| `Initialize` | ∅ | ∅ | Initializes the ADT |
| `Append` | ITEMSET | ∅ | Records a new item into the counts |
| `Expire` | ITEMSET | ∅ | Removes an item from the counts |
| `QueryHeaviest` | $k$: INTEGER | ARRAY[$k$] | Returns the heaviest-$k$ ITEMSETs |
| `QueryLightest` | $k$: INTEGER | ARRAY[$k$] | Returns the lightest-$k$ ITEMSETs |

**Table 1.** The HL-HITTERS Abstract Data Type

### 2.1 Building Blocks

To implement the data structure we use common basic building blocks. More specifically, we use exactly one array of fixed size, one doubly linked list and one hash table. With each of these data structures we only use the constant time operations. Thus, for example, we never iterate over the nodes of the linked list to reach a sought entry, rather we keep references to the node itself. We will proceed by describing exactly which operations will be used on each data structure and its time complexity.

**Array** The array must be of size $Q$, the same as the size of the window, and its size remains constant during the execution of the algorithm. We only perform the operations `Get` and `Set` on the array, which execute in constant time. The elements of the array are never iterated over.

In the implementation for our experiments we used the standard vector provided by the C++ STL STD::VECTOR class.

**Doubly-linked list** The linked list starts out empty and as the algorithm executes nodes are added and removed. We only use the *Head* and *Tail* fields of the doubly-linked list to access the respective nodes in constant time. As far as the inserts and deletes are concerned, they are always executed with respect to a reference node and as such are constant time as well. To be more specific, `InsertBefore` and `InsertAfter` require two arguments: the new node to insert and a reference node before or after which to insert the new node. Similarly, `Delete` requires a direct reference to the node to delete. Furthermore, the maximum number of nodes is known a priori to be $Q$, and thus we can eliminate the overhead of dynamic memory allocation for the nodes by using a preallocated node pool.

In the implementation for our experiments we used the standard doubly-linked list provided by the C++ STL STD::LIST class.

**Hash-table** In the HL-HITTERS data structure the id of each itemset with at least one item in the window, is stored in a dynamic dictionary. A hash-table is used to implement the dynamic dictionary. Hashing is commonly assumed to require $O(1)$ amortized time for the operations `Get`, `Set` and `Delete` or at least for one of these operations. However, there are at least two examples of hashing schemes which achieve worst case $O(1)$ time with high probability (whp): the early work of [5] and the recent algorithm of [2]. Consequently, we can assume that an efficient, $O(1)$ hashing scheme can be used in the HL-HITTERS data structure.

There is an additional reason why we can assume $O(1)$ time for our hashing scheme. Given that our original motivation were router queues, we can assume that the maximum size of a window does not typically exceed 1000 items (packets in this case). The most common values are a few hundred items. This fact admits us the luxury to run the hashing data structure with a very low load factor. For example, even a hash table with 1 million entries would not be a significant cost for a modern router.

Consider the following naive approach with chained hashing using a uniform hashing function with $n$ hash table entries, $m \ll n = cm$ packets, and $k$, the constant upper bound on the number of collisions. The probability $\rho$ of experiencing more than $k$ collisions in any of the $n$ table entries is

$$\rho \leq n \cdot \binom{m}{k+1} \left(\frac{1}{n}\right)^{k+1} \leq n \left(\frac{e}{k+1}\right) \left(\frac{1}{c}\right)^{k+1}. \tag{1}$$

For $n = 10^6$, $m = 10^3$ and $k = 10$ the first inequality gives that $\rho \leq 2.38 \times 10^{-35}$. Consider now a router which serves $10^9$ packets per second (a bit unrealistic today but lets allow for future enhancements) and operates continuously for 20 years. This router can serve not more than $Z = 10^9 \times 60 \times 60 \times 24 \times 366 \times 20 \leq 6.34 \times 10^{17}$ packets during its lifetime. Even if we consider the case where every one of these $Z$ packets is unique, i.e., the router never receives two packets from the same flow and thus maximizes the potential for collisions to appear, the probability of a "bad" collision event occurring during its lifetime is $\rho * Z \leq 2.38 \times 10^{-35} \times 6.34 \times 10^{17} = 1.51 \times 10^{-17}$. This probability is thus practically negligible. Consequently, even the naive approach seems to meet the requirements for a router. In addition to this naive implementation there are many, very efficient, hashing schemes which will perform much better. The question of which of the published hashing schemes offers the optimal trade-off between space redundancy and worst case bounds could be an interesting problem to investigate. However, for our purposes, any lightweight hashing scheme will be sufficient if sufficient memory is provided. Moreover, for our main motivation application, special hardware-based memory is available in many routers which can achieve de-amortized $O(1)$ performance [10].

Based on the above arguments, we plausibly assume that we can employ an efficient $O(1)$ whp hashing scheme for our data structure in a modern network router. Additionally, we believe that the arguments used for the router case can apply to other applications of window-based heaviest and lightest hitter problems. In the implementation used for the experiments of this work, we used chained hashing provided by the C++ BOOST::UNORDERED_MAP class[1].

## 2.2 Data Structure

We now proceed to describe how the data structure is composed out of the basic building blocks. An overview of the layout used is presented in Figure 1.

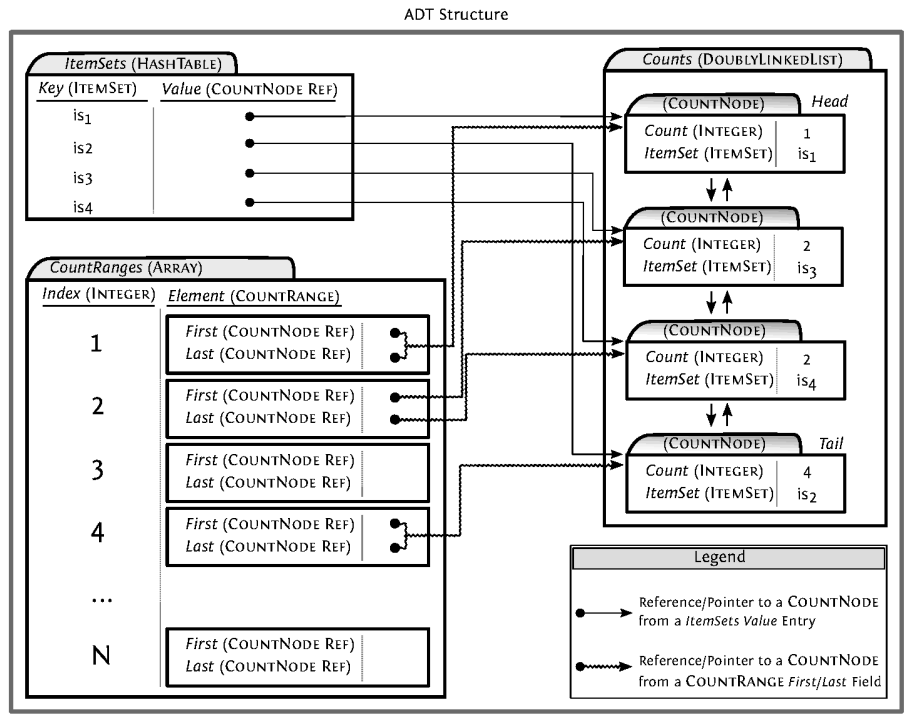

Fig. 1. The ADT's structure.

Before proceeding with the description of the data structure further, we need to describe two types of simple record-like structures which are used:

- COUNTNODE, which is the type of the list node used in the doubly-linked list. The data stored (besides the *Previous* and *Next* fields) is an integer named *Count* and the identifier of an ITEMSET named *ItemSet*.
- COUNTRANGE, which has two fields, named *First* and *Last*, both of which are references to a doubly linked list node of type COUNTNODE. This structure is meant to store the endpoints of a sub-range of the *Counts* DOUBLYLINKEDLIST. To support this, it supports two simple operations: Insert (a new node in range) and Remove an existing node from the range. Both are $O(1)$ operations as they manipulate only the *First* and *Last* fields and do not iterate over the nodes in the range.

**Layout of the Data Structure** Itemsets that have no items in the window, i.e., a count of zero, will not have any entries in any of the data structures. Conversely, each itemset which has at least one item in the window, i.e., a count $\geq 1$, will have one entry in the *ItemSets* HASHTABLE. Additionally, for each itemset, there will exist one node of type COUNTNODE in the *Counts* DOUBLYLINKEDLIST, with a *Count* field corresponding to its exact count of items in the window. Finally, for each group of itemsets which have the same item count there will be one entry in the *Ranges* ARRAY, in the position of the array which is equal to the itemset group's count.

### 2.3 Algorithms

We now present the operations which are supported by the data structure using pseudo-code and describe their operation and computational complexity in detail.

---

**Algorithm 1** The `Initialize` operation on HL-HITTERS

1: **procedure** `Initialize`
2:     *ItemSets* ← **new** HASHTABLE
3:     *Counts* ← **new** DOUBLYLINKEDLIST
4:     *Ranges* ← **new** ARRAY
5: **end procedure**

---

**Initialization** The `Initialize` operation is shown in Algorithm 1. While its functionality is simply to initialize the *ItemSets* hash table, the *Counts* doubly linked list and the *Ranges* array, it is useful nevertheless to illustrate that initialization is straightforward and that only memory allocations are performed.

**Append** In Algorithm 2 we present the `Append` operation. It receives the itemset of the item which is to be appended as a parameter. The itemset is looked up in the *ItemSets* hash table. If it is found, then the itemset is already being counted, i.e., has other items in the window, and therefore its count must be increased by one. If not, then it is a new itemset, i.e., it has no other items in the window, and thus must be recorded with a count of one.

For the case of being already counted, only the *Counts* and the *Ranges* structures will be modified. The idea is to move the count node corresponding to the itemset to the position in the *Counts* linked list where it will be the first linked list node with the new count. In order to do this, the count node of the itemset is looked up via the `Get` operation on the hash table and a reference to it is stored in $cn$. Before removing the $cn$ node from the list, the position in the linked list where it will be moved to is recorded in $cn'$, with help from the *Ranges Last* field. This will point to the immediately next linked list node after the last node with the old count. Subsequently, the count node $cn$ is removed from the linked list and the corresponding *Ranges* count range entry is updated with the `Remove` operation.

**Algorithm 2** The `Append` operation on HL-HITTERS
---
1: **procedure** Append(*itemset*: ITEMSET)
2:     $cn \leftarrow cn' \leftarrow$ **null**
3:     **if** *itemset* $\in$ *ItemSets* **then**
4:         $cn \leftarrow$ *ItemSets*.Get(key:*itemset*)
5:         $cn' \leftarrow$ *Ranges*.Get(index:*cn.Count*).*Last.Next*
6:         *Ranges*.Remove(node:*cn*)
7:         *Counts*.Remove(node:*cn*)
8:         *cn.Count* $\leftarrow$ *cn.Count* $+ 1$
9:         *Counts*.InsertBefore(beforenode:*cn'*, node:*cn*)
10:        *Ranges*.Insert(node:*cn*)
11:     **else**
12:        $cn \leftarrow$ **new** COUNTNODE(ItemSet:*itemset*, Count:1)
13:        *Counts*.InsertBefore(beforenode:*Counts.Head*, node:*cn*)
14:        *Ranges*.Insert(node:*cn*)
15:        *ItemSets*.Set(key:*itemset*, value:*cn*)
16:     **end if**
17: **end procedure**

Finally, the *cn* node is inserted in the linked list before the *cn'* node and the new *Ranges* count node entry is updated to include it.

For the case of not being already counted, all of the structures will be modified. A new count node will be created to hold the count for the new itemset. Since allocating a new object on the heap may not be $O(1)$, we can take advantage of the fact that the maximum number of itemsets is $Q$, as explained in Section 2.1, and as such we can just take out a preallocated count node out of a preallocated pool in $O(1)$. This node is then inserted in the position of the *Counts* linked list indicated by the *First* field in the first count range entry of the *Ranges* array and then it is recoded in the same count range entry. Finally, the *itemset* hash table is updated by creating an entry that maps the new itemset to the count node which was created previously using the `Set` operation.

**Expire** In Algorithm 3 we present the `Expire` operation. It receives the itemset of the item which is to be removed as a parameter. The itemset is looked up in the *ItemSets* hash table via the `Get` operation and the reference to the count node in the *Counts* linked list representing it is stored in *cn*.

Since the count of the itemset will be decremented by one, we need to move the *cn* count node to the position in the *Counts* linked list where it will be the first linked list node with the new (old minus one) count. Similarly to the `Append` operation, before removing the *cn* node from the list, the position in the linked list where it will be moved to is recorded in *cn'*, with help from the *Ranges First* field. This will point to the immediately previous linked list node after the first node with the old count. Subsequently, the count node *cn* is removed from the linked list and the corresponding *Ranges* count range entry is updated with the `Remove`

**Algorithm 3** The Expire operation on HL-HITTERS

```
 1: procedure Expire(itemset: ITEMSET)
 2:     cn″ ← null
 3:     cn ← ItemSets.Get(key:itemset)
 4:     cn′ ← Ranges.Get(index:cn.Count).First.Previous
 5:     Ranges.Remove(node:cn)
 6:     Counts.Remove(node:cn)
 7:     cn.Count ← cn.Count - 1
 8:     if cn.Count ≥ 1 then
 9:         if cn′ ≠ null and cn′.Count = cn.Count then
10:             cn″ ← Ranges.Get(index:cn′.Count).First
11:             Counts.InsertBefore(beforenode:cn″, node:cn)
12:         else
13:             Counts.InsertAfter(afternode:cn′, node:cn)
14:         end if
15:         Ranges.Insert(node:cn)
16:     else
17:         delete cn
18:         ItemSets.Delete(key:itemset)
19:     end if
20: end procedure
```

operation. The count node *Count* field is decremented by one. If the count has not reached zero a check is made to see whether the position to be moved is valid:

- The reference in $cn'$ must be not null, which would indicate that the previous count range was the first in the linked list, and
- the count of the $cn'$ referenced node must be the same as the new count of the moving node, i.e., the target count node must belong to the correct count range.

If this check succeeds, the new corresponding *Ranges* count range entry is fetched with the Get operation. Its *First* field is set as the new $cn''$ insertion position. Afterwards the moving node is inserted there. If the check fails, then there is no COUNTRANGE entry in the *Ranges* array corresponding to the new count and the count node is inserted right where the original $cn'$ reference pointed to.

In both cases, the moving count node will be inserted in the *Ranges* entry with the new count using the Insert operation.

If the new count after decrementing by one is zero, the count node is deleted. If a preallocated pool was used it is returned to the pool in $O(1)$. Finally, the *itemset* hash table is updated by deleting the entry that maps the itemset to the count node which was previously deleted.

**Query** In Algorithm 4 we present the QueryHeaviest and the QueryLightest operations simultaneously. The basic algorithm is the same; only the start of the iteration and its direction is different. In the algorithm, the left side of the ◀|▶

**Algorithm 4** The `Query Heaviest ◄|► Lightest` operation on HL-HITTERS
─────────────────────────────────────────────
 1: **function** QueryHeaviest(*k*: INTEGER)
 2:    *results* ← **new** ARRAY[*k*]
 3:    *cn* ← *Counts.Tail* ◄|► *Counts.Head*
 4:    *i* ← 1
 5:    **while** *i*≤*k* **and** *cn*≠**null do**
 6:       *results[i]* ← *cn.ItemSet*
 7:       *cn* ← *cn.Previous* ◄|► *cn.Next*
 8:       *i* ← *i*+1
 9:    **end while**
10:    **return** *results*
11: **end function**
─────────────────────────────────────────────

symbol corresponds to the `QueryHeaviest` operation while the right side to the `QueryLightest` operation.

The algorithm receives the threshold $k$ as a parameter. Initially, a new *results* array of size $k$ is created to hold the results. In some cases, there may be less than $k$ itemsets available, therefore a number of positions at the end of the array will have null entries.

The count node reference *cn* is set to point to the last (for `QueryHeaviest`) or the first (for `QueryLightest`) node in the *Counts* linked list via its *Head* or *Tail* fields. Afterwards, an iteration is performed up to $k$ times. In each step, the current itemset stored in the node referenced by *cn* is stored in the current (the $i$-th) index of the array. Finally, the result is returned.

The whole operation makes up to $k$ iterations, at each one adding a different itemset to the result. This makes this operation have a time complexity of $O(k)$ and as such is constant time as well. The operation of the query algorithm can easily be extended without changing the computational complexity to also return the actual count of each itemset along with each itemset. In addition it is possible instead of specifying a $k$ parameter to return all the itemsets with the highest/lowest count. To implement this, retrieve the *Tail*/*Head* count node of *Counts*, get the highest/lowest count, access the *Ranges* entry corresponding to that count and get the range of count nodes between the *First* and *Last* fields with the max/min count. This algorithm's computational complexity will depend on the number of itemsets which will be the max/min count. As it is possible to have $Q$ itemsets each with a count of one, this algorithm will have a worst case complexity of $O(Q)$. However, in practice in many applications this will seldom be the case. Another extension would be to return the heaviest-$\theta$/lightest-$\theta$ hitters, where $\theta$ is relative, expressed as a proportion of the window size (e.g. $\theta$= 10%). However, here the `QueryHeaviest` and the `QueryLightest` operations will have different complexities. Since there is an upper bound on the number of itemsets which can have a frequency more than or equal to $\theta$ equal to $1/\theta$, one can just execute `QueryHeaviest` with $k = 1/\theta$ and the complexity will be as originally $O(k)$. However, no such bound exists for the `QueryLightest` case, and therefore its worst case complexity will be $O(Q)$.

Finally, if one is willing to accept an $O(Q)$ worst case complexity it is possible to create cumulative versions of both the original and the relative version of the query operations, where the $k$ or $\theta$ parameters denote the cumulative count or proportion of the window. This would return the first itemset whose counts together add up to the specified threshold.

### 2.4 Space Complexity

The space complexity of the HL-HITTERS data structure can be fully derived and is exclusively dependent on the maximum window size $Q$. The *ItemSets* hash table contains a maximum of $Q$ entries, the *Ranges* array has a constant size of $Q$ entries and the *Counts* doubly linked list contains a maximum of $Q$ count nodes. It follows that the space complexity of the whole HL-HITTERS data structure is $O(Q)$.

## 3 Experimental Evaluation

It is clear from the previous analysis that the computational complexity of the algorithms presented is overall constant time whp. However, this does not guarantee an acceptable level of performance if in practice the constant time required is too high. We have created a router-like scenario, and have performed experiments to gauge the actual performance of the proposed algorithms. We have to note that, to our knowledge, there exists no other algorithm for calculating the heaviest-$k$ hitters exactly, which also provides close to constant time performance. Therefore, we have implemented a naive but efficient as far as possible algorithm to find the heaviest-$k$ hitter. This algorithm, each time the heaviest hitter is requested, creates a hash-table, and records within it the counts for each itemset. As it does this, it keeps track of the running heaviest hitter. However, it is clear that this algorithm has an $O(Q)$ time complexity. As each item arrives for processing, it is recorded in the counts and immediately afterwards the heaviest hitter is queried. This represents the worst case scenario, where the query operation is performed at each time step. Furthermore, in the experiments performed, we restricted ourselves to finding the top heaviest hitter only, i.e., $k=1$, in order not to significantly disadvantage the direct counting algorithm.

### 3.1 Experiment Setup

The implementation has been performed using C++, with standard C++ versions of the building blocks, as described in section 2.1. We used the G++ compiler with all the optimizations enabled ($-O3$) for our specific architecture. The experiments were executed on an Intel Quad Core Q9300 processor with $4GB$ of main memory, using one dedicated core for the execution of the experiments. The operating system used was Arch Linux, with the 2.6.36 version kernel. For each result point 10 identical sequential executions of the experiment were performed to remove any bias.

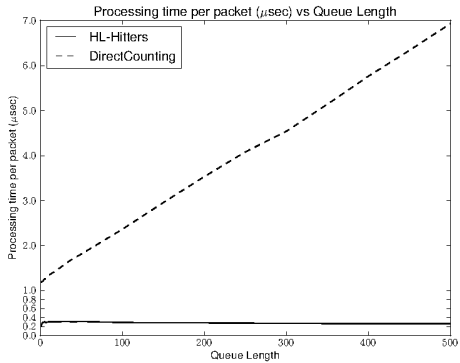
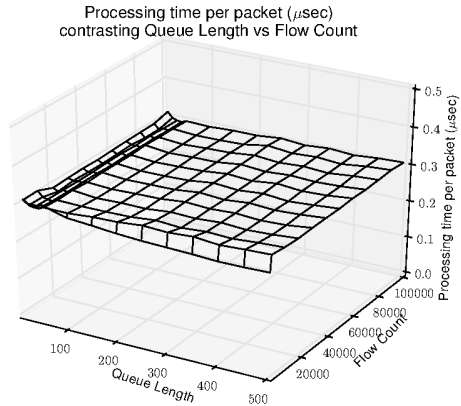

**Fig. 2.** Performance of HL-HITTERS vs direct counting for different $Q$ queue lengths. Measured in mean processing time per packet (shown in $\mu s$).

**Fig. 3.** Performance of HL-HITTERS for different $Q$ queue lengths and number of flows. Measured in mean processing time per packet (shown in $\mu s$).

### 3.2 Results

The results of the experiments are displayed in Figures 2 and 3. Figure 2 shows the results of the comparison between HL-HITTERS and the direct counting algorithm. As predicted, the direct counting algorithm exhibits performance which is a linear function of $Q$. At the same time, HL-HITTERS maintains constant performance in accordance to $O(1)$. It is noteworthy to examine the absolute numbers as well. The HL-HITTERS algorithm has an approximate processing time per packet of $0.3\mu s$. This means that despite using general purpose building blocks and no hardware-based content addressable memory or specialized CPUs, we can process approximately 3.3 million packets per second using our implementation. According to [11] IP packet sizes vary between $40 bytes$ and $1500 bytes$, with strong polarization tendencies. Given those values, we can achieve a throughput between $1 Gbit/sec$ and $40 Gbit/sec$. We stress the fact that this performance is achievable without any specialized hardware as would typically exist in an Internet backbone router. Furthermore, preliminary performance profiling has shown that approximately 45% of the processing time is spent on the hash-table operations and approximately 20% on the doubly linked list operations. Since both would heavily benefit from optimizations on a hardware router, we are confident that significantly higher performance is attainable under such conditions.

The data plot in Figure 3 presents how the per-packet processing time varies with both the length of the queue $Q$ and the number of different flows in the system. This experiment aims to examine whether an increase in the queue length or an increase in the number of flows (leading to more collisions in the hash-table) will impact performance. To the extent that we could see, neither of those factors impacted performance.

## 4  Discussion

Our work on the problem of the heaviest-$k$ and lightest-$k$ hitters in a sliding-window data stream has resulted in a relatively simple data structure and an efficient set of algorithms for its operations. These in tandem allow us to achieve constant time updates and queries, something which, to our knowledge, has been achieved for the first time. Moreover, and for some applications more importantly, the performance of this scheme has been verified to be high enough to be used in practical applications. Lastly, the fact that we haven't presented a highly optimized and hardware assisted implementation allows us to predict much better performance in practical application where these additional enhancements would be pursued. As we have described, the performance of this algorithms is dependent only on the available memory, and especially the memory for the hash-table.

An interesting idea beyond these results would be to extend this mechanism to incorporate the size of the packets as well, not only their number. This would allow us to make decisions based on the quantity of data that an itemset is responsible for, rather than how many items it is generating.

**Acknowledgements** We would like to thank Dimitrios Fotakis for our insightful discussions on efficient hashing.